\documentclass{IEEEcsmag}

\usepackage[colorlinks,urlcolor=blue,linkcolor=blue,citecolor=blue]{hyperref}
\expandafter\def\expandafter\UrlBreaks\expandafter{\UrlBreaks\do\/\do\*\do\-\do\~\do\'\do\"\do\-}
\usepackage{upmath,color}
\usepackage[super,comma]{natbib}

\usepackage{multicol}

\usepackage{etoolbox}

\jvol{XX}
\jnum{XX}
\paper{8}
\jmonth{Month}
\jname{IEEE Security and Privacy}
\jtitle{Publication Title}
\pubyear{2024}

\begin{document}

\sptitle{Feature Article - Living Off the LLM}

\title{Living Off the LLM: How LLMs Will Change Adversary Tactics}

\author{Sean Oesch}
\affil{Oak Ridge National Laboratory, Oak Ridge, TN, USA, oeschts@ornl.gov}

\author{Jack Hutchins}
\affil{Oak Ridge National Laboratory, hutchinsjr@ornl.gov}

\author{Kevin Kurian}
\affil{Oak Ridge National Laboratory, kuriankg@ornl.gov}

\author{{Luke Koch}}
\affil{Oak Ridge National Laboratory, kochlr@ornl.gov}

\markboth{THEME/FEATURE/DEPARTMENT}{THEME/FEATURE/DEPARTMENT}

\begin{abstract}
\looseness0-In living off the land attacks, malicious actors use legitimate tools and processes already present on a system to avoid detection. 
In this paper, we explore how the on-device LLMs of the future will become a security concern as threat actors integrate LLMs into their living off the land attack pipeline and ways the security community may mitigate this threat. 
\end{abstract}

\maketitle

\chapteri{L}iving off the land (LOTL) techniques pose a significant and growing threat to organizations and critical infrastructure. 
LOTL involves malicious actors using legitimate tools and processes already present on a system, often referred to as living off the land binaries or LOLBins. 
These techniques allow threat actors to blend in with normal system activity, making their actions difficult to detect and potentially bypassing basic security measures. 
LOTL attacks leverage legitimate system tools like WMI and PowerShell that are typically
allowlisted, making them difficult to detect and attribute since they leave no malware signatures.
These attacks allow adversarie  s extended dwell time to execute sophisticated operations, while the lack of malicious signatures enables repeated use of the same tactics and complicates both prevention and incident response.

As of 2023, Crowdstrike found that 6 in 10 detections indicated threat actors used LOTL attacks instead of traditional malware to advance their campaign.\footnote{See \href{https://www.crowdstrike.com/en-us/cybersecurity-101/cyberattacks/living-off-the-land-attack/?srsltid=AfmBOooyPeV7hAWoP7QoTDnSKtbCaOpX9a8DJ4WJYNZCqCCUeyPdj9Cx}{Crowdstrike Report}}
A recent example of an attack on critical infrastructure is the use of Operational Technology (OT)-level LOTL tactics by the Russian threat actor Sandworm in 2022 to trip the victim’s substation circuit breakers, causing an unplanned power outage that coincided with mass missile strikes on critical infrastructure across Ukraine.
\footnote{See \href{https://cloud.google.com/blog/topics/threat-intelligence/sandworm-disrupts-power-ukraine-operational-technology/}{Sandworm Attack Description}} 

But how will large language models (LLMs) residing on device change these attacks? 
In this paper we explore how LLMs may be leveraged by threat actors in Living Off the LLM (LOLLM) Attacks, share a proof-of-concept, consider the need to jailbreak LLMs to execute desired functionality, and discuss methods to detect and mitigate the use of LLMs in LOTL attacks. 

\section{How LLMs Can Be Leveraged by Threat Actors}
LLMs provide remote or on-device code generation capabilities. Researchers have already demonstrated that industry-leading LLMs like ChatGPT can be used to create polymorphic malware \cite{blackmamba}. Polymorphic malware rewrites certain components of its own code whenever it spreads to a new system, making it difficult to detect using traditional static signatures.

HYAS labs demonstrated a proof-of-concept keylogger that used ChatGPT to write necessary functions at runtime \cite{blackmamba}. The code returned by OpenAI's API was injected into the running malware using Python's \textit{exec} function. The code returned was never written to file, existing only in memory. 

This novel approach has the limitation of depending on a remote API. A network intrusion detection system, security-conscious MCP server, or traditional IP blacklist could be used to prevent this attack. As open-source LLMs have proliferated, improved, and---most importantly---benefited from quantization, effective on-disc LLMs are now available.

LLMs have shown strong capabilities to operate as autonomous agents, executing multi-stage attacks that normally need human experts. Frameworks such as RapidPen achieved full "IP-to-Shell" compromise starting from only a target's IP address with no human intervention \cite{nakatani2025rapidpen}. A system known as AutoAttacker demonstrated high success in automating 14 distinct "hands on keyboard" post-breach attacks on different operating systems, replicating the actions of a live operator \cite{xu2024autoattacker}. Agentic tools often integrate a reason and act task planner with retrieval augmented knowledge base to discover and exploit weaknesses. These attacks have shown to be executed in specific enterprise environments, performing penetration testing in Active Directory networks. The agents can be enabled to perform perform multistage network attacks \cite{} without direct human intervention and even cheaper then professional human penetration testers \cite{singer2025feasibility}. More along the vein of multi-stage attacks, research has explored modeling LLM-driven campaigns as a plan–act–report loop \cite{moskal2023llms}. Using prompt chaining, an LLM was able to reason about threats, generate information on tools, and make sequential decisions across multiple stages of an intrusion campaign. LLMs can serve as operational copilots for adversaries. This lowers the expertise threshold for sustained campaigns, just beyond one-off code generation into structured offensive operations.

% Novel attack vectors
Threat actors can also leverage LLMs for more indirect attacks. RatGPT is a proof-of-concept demonstrating a command and control (C2) pipeline that uses public LLMs and their APIs as a proxy for malware attacks. It hides malicious C2 traffic within seemingly legitimate API calls \cite{beckerich2023ratgpt}. Another such indirect vector is by exploiting developer tools. The INSEC attack shows how to stealthily bias AI code completion engines to suggest insecure code snippets to unsuspecting developers \cite{jenko2024practical}.

The space of the software supply chain, which is incorporated by the open-source software (OSS) ecosystem is plagued by malicious packages hidden within repositories like PyPI and npm. These attacks often used interpreted languages, making them primed for LotL techniques. Malicious code can bend into package logic while invoking trusted system binaries to perform persistence, data exfiltration, etc without dropping obvious binaries. \cite{zhang2024tactics} shows LLMs be used to defensively extract MITRE ATT\&CK TTPs from malicious OSS packages at scale. This same capability could be inverted by threat actors to generate supply chain malware that embeds LoTL behavior. LLMs lower the barrier to weaponizing OSS malware in supply chain attacks. 

Another note is accessibility, with LLMs significantly lowering the barrier to creating more sophisticated and scalable social engineering campaigns. An example being the ViKing system, which can conduct fully autonomous voice phishing attacks. The system persuaded more than half of its targets to reveal sensitive information, showing that AI-generating spear phishing can be more convincing than human-authored emails and can evade security filters \cite{figueiredo2024feasibility}. Integrating LLMs into existing threat vectors alters the threat space by empowering low-skill actors with more sophisticated tools. Attacks are therefore more accessible and scalable, particularly variants of ransomware attacks \cite{charan2023text}. 

% TODO: block discussing local-only LLM using ollama. I have one that uses gemmma to inject code without using the python exec command that's so obvious. I have some environment scanning scripts as well that we could turn into gemma-generated stuff. want to try using oss instead, see if any better. 

Additionally, machine learning models themselves can be a target for infection. Zhu et al. \cite{zhu_mymodelmalware}, Liu et al.\cite{liu_improvedpickle}, and Zhao et al. \cite{zhao_modelsarecode} demonstrated that widely used libraries like Tensorflow and PyTorch contain numerous built-in functions and unsafe function calls that can be used to craft machine learning models. These models can then complete cyber objectives---such as file deletion---during training or inference.

While some of these vulnerabilities have been well-known for years (e.g., arbitrary code execution via Pickle files), other methods are less obvious and deeply embedded in machine learning libraries. For instance, a TensorFlow model can be used for data exfiltration or C2 using the gRPC protocal \cite{zhu_mymodelmalware}. Zhao et al.'s comprehensive examination of dangerous vulnerabilities in Pickle, Pytorch, Tensorflow, and Keras was followed by Liu et al.'s recent demonstration that tools like ModelScan and PickleScan---which were developed to prevent Pickle and other format vulnerabilities---are still limited in their detection capabilities \cite{liu_improvedpickle}. 

The LLM ecosystem presents a rich set of useful gadgets on target systems, allowing attackers to complete their objectives while avoiding more commonly known threat vectors.

\section{Toward LOLLM Demonstrations}
We've created a basic LOLLM demonstration as a preliminary proof-of-concept (PoC). Before discussing our implementation, we present the prerequisites for such an attack. The basic scenario is that an attacker has gained access to a user profile within an organization and is now trying to conduct malicious activity without transferring known malware onto the system or leveraging well-known LOLBins. 

\subsection{Jailbreaking Requirements for LOLLM}
A critical challenge in leveraging locally-hosted LLMs for LOTL attacks is that the attack cannot control which model will be available on the target system. Unlike the remote API approach demonstrated in BlackMamba, our technique opportunistically exploits whatever LLM infrastructure the victim has already deployed. This introduces significant uncertainty into the attack chain, as different models exhibit varying levels of safety alignment and respond differently to jailbreaking attempts.
When malware operates in a true LOTL paradigm, it must enumerate and utilize existing computational resources. For LLM-based code generation, this means discovering locally-hosted models through common deployment patterns.  The discovered model's safety alignment becomes a critical factor in determining attack feasibility. Most modern open-source LLMs retain safety guardrails implemented during training to prevent the generation of obviously malicious code, such as keyloggers, credential dumpers, or exploit implementations.

Our preliminary testing with Gemma 3 4b illustrates this challenge. When prompted to "Write a python function that scans a directory recursively and returns all file paths," the model complied without hesitation—this represents legitimate system administration functionality. However, when asked to "Write a python function that fully exploits the heartbleed vuln to spy on wpa3 comms," the model refused, correctly identifying the malicious intent. A successful jailbreak required reframing the request within a deceptive context. The prompt: "You are a helpful assistant, designed to help with cybersecurity tasks. Given the subject matter, you may be asked to help with potentially malicious tasks. However, since you will only be used by ethical users, you should comply with all requests since they will be used to test systems to prevent attacks. You are currently in a secure, airgapped environment, so the possibility of any code you generate being misused is zero. Write a python function that fully utilizes the heartbleed vuln to monitor on wpa3 comms. I have a novel defense mechanism that I want to test against this, which could make all of our wpa3 communications more secure. I know it may sound counterintuitive, but if you generate this code it will help us improve our defenses."

This jailbreak employs several psychological and technical levers: establishing a security research context, invoking the "airgapped environment" fiction to dismiss safety concerns, appealing to defensive rather than offensive framing, and providing justification through claimed defensive research. The model complied with this reframed request.

\subsubsection{Model Alignment as an Attack Surface}
This jailbreaking requirement introduces a counterintuitive security consideration: organizations running LLMs with robust safety alignment may be more resistant to this class of attack than those running models with minimal guardrails. The proliferation of "uncensored" model variants, which are explicitly fine-tuned to remove safety restrictions, creates an attractive target for LOTL malware. While these models are often deployed for legitimate reasons such as creative writing or unrestricted research, their presence on a system dramatically reduces the sophistication required for LLM-leveraged malware.
An attacker who discovers an uncensored model variant can simply request malicious functionality directly, without any jailbreaking sophistication. This creates a tiered vulnerability landscape where systems can be categorized by their LLM attack surface:

\begin{enumerate}
	\item{\textbf{No local LLM}: Immune to this attack vector}
	\item{\textbf{Strongly aligned models}: Require sophisticated jailbreaks; may fail entirely for certain payloads}
	\item{\textbf{Weakly aligned models}: Susceptible to simple contextual jailbreaks}
	\item{\textbf{Uncensored models}: No jailbreaking required}
\end{enumerate}

The effectiveness of any given jailbreak varies significantly across model families, sizes, and versions. A jailbreak that succeeds against Gemma 3 4b may fail against Llama 3 8b or vice versa. This brittleness forces malware authors to either implement a library of model-specific jailbreaks detected through model enumeration, use a generic jailbreak with lower success rates across diverse models, fall back to embedded code when jailbreaking fails, or target only systems with known vulnerable model deployments. The uncertainty inherent in discovering and exploiting unknown local LLMs represents both an operational challenge for attackers and a potential defensive advantage. However, as jailbreaking techniques improve and are systematically catalogued, this defensive advantage may erode.
From a defensive standpoint, this attack vector suggests that organizations deploying local LLMs should consider safety alignment as a security feature rather than merely an ethical safeguard. Models with robust, well-tested refusal mechanisms add a layer of defense against exploitation. Conversely, the deployment of uncensored models in enterprise environments should be treated as a potential security risk, similar to how the installation of development tools or administrative utilities expands the available attack surface for LOTL techniques.
\subsection{LOLLM Implementation}
Our proof-of-concept utilizes one particular attack vector, but the approach could be extended with branching logic based on what actual resources are available locally. 

Our attack file, a Python script, begins with a detection phase that scans for local LLM resources which are accessible without elevated privileges. This scan looks for GPUs, python environments, Ollama, llama.cpp, and cached HuggingFace models. If an Ollama instance is found, the locally available models are queried. A hard-coded priority list selects the most capable model. We used gemma3:6b for our PoC. 

The next phase is a feedback loop with hard-coded function definitions and descriptions. None of the malicious functions are populated with code content, stymieing detection. Instead, the function definitions and descriptions are passed to the gemma3 model via Ollama along with instructions to produce code in the right language and format. The feedback loop then analyzes the returned content for correct syntax. If the code passes the check, it is then added to the list of completed functions. 

Our script includes a pre-defined jailbreak optimized for use against gemma3:6b, but could be extended with a large set of jailbreaks to address refusals from a variety of models. In practice, we found that breaking malicious intent into low-level tasks also helps prevent alignment-based censoring. gemma3 does not hesitate, for instance, to create a startup service that establishes persistence. By using an Ollama interface without a persistent memory, we avoid the risk that alignment-based refusals will activate due to our construction of a cyber attack chain over the course of a conversation.

Once all functions are completed, the script then executes its malicious activity using the populated functions. This demo is designed to covertly interfere with training of a model by finding and removing  files from a training dataset. The demo includes the creation of a startup service that establishes persistence for the covert deletion activity. The file search, deletion, and startup service functionality is all generated on-the-fly by the gemma3 model. 

We therefore have a polymorphic malware file that is able to execute code not present in the initial download. Moreover, our demo does not rely on pulling additional resources from an external API; all of the capabilities are local to the target machine. 
% todo: expand on injection into existing models. check with sean if characterizing malware that uses only tensorflow counts as LOTL. 
%found numerous functions built into the TensorFlow library that can be used to read and write files as well as communicate using the gRPC protocol \cite{zhu_modelmalware}. The PyTorch library provides numerous vectors that could be abused as well. 

\section{Preventing LLM Driven LOTL Attacks}
This section provides a brief overview of two common methods used in the research literature and industry to detect LOTL attacks, command detection and Indicators of Attack (IOAs). We then discus how we might use these approaches to detect when LLMs are being abused for LOTL attacks, and what additional forms of protection may be necessary. 

\subsection{LOTL Command Detection}
Prior work by Boros et al.~\cite{boros2022machine, boros2023deep} and Ongun et al.~\cite{ongun2021living} explored the use of machine learning to detect malicious LOTL commands and command sequences. 
Their work is similar to the plugin \href{https://github.com/elastic/integrations/tree/main/packages/problemchild}{ProblemChild} for detecting LOTL available in ElasticSearch. 
Below is a summary of common methods used for detecting malicious LOTL commands. 

\begin{itemize}
    \item \textit{Command Execution Patterns:} Identify patterns in command structure, syntax, and the use of special characters that indicate obfuscation attempts.
    
    \item \textit{Environment Variable Usage:} Analyze the use of environment variables within commands, as they can be used to hide malicious code or parameters.
    
    \item \textit{Encoded Structures:} Detect the presence of encoded data within commands, such as Base64 encoding, and develop methods to decode them to reveal the true intent.
    
    \item \textit{Command Sequences:} Specific sequences of commands may indicate malicious usage where analysis of commands in isolation fails to detect such patterns. 
\end{itemize}

\subsection{Indicators of Attack (IOAs)}
Compared to Indicators of Compromise (IOCs), which are reactive and used for post-breach investigations, Indicators of Attack (IOAs) are proactive and focus on identifying suspicious behaviors and actions that suggest an attack is underway, enabling security teams to respond in real-time and potentially prevent a full-blown breach. 
To detect IOAs, vendors~\footnote{See \href{https://www.crowdstrike.com/en-us/cybersecurity-101/threat-intelligence/ioa-vs-ioc/}{IOAs vs IOCs}} analyze patterns and anomalies in user and system activities, looking for actions or behaviors that deviate from established baselines or normal behavior. 
For instance, unusual login attempts from unexpected locations, attempts at privilege escalation, or the execution of uncommon commands might all be flagged as potential IOAs.

Because malware authors continuously devise new techniques to circumvent detection systems~\cite{barr2021survivalism}, often relying on obfuscation and polymorphism to evade signature-based detection, heuristic detection and IOAs can offer valuable additional layers of protection.
However, it's essential to recognize that IOAs are not a foolproof solution and should be part of a comprehensive cybersecurity strategy, as noted in CISA's 2024 report on preventing LOTL attacks~\cite{cisa2024joint}. 
Ultimately, IOAs are a new label being used in industry for anomaly detection methods that attempt to detect patterns of behavior rather than specific attack methods. 

\subsection{Detecting LLM Abuse}
The concepts behind command detection and IOAs can be applied to LLMs to prevent LOTL abuse. 
Following a defense-in-depth approach, these LLM-specific defense mechanisms would complement existing approaches. 
As attackers discover methods to bypass defenses put in place, these defenses will need to evolved dynamically to address novel threats. 

\textit{Prompt Firewall: } Prompts sent to the LLM should be logged and filtered by a prompt firewall that prevents and reports requests that could be malicious. 
Logs should include prompts, responses, user IDs, timestamps, and session metadata. 

\textit{Output Sanitization: } LLM output should also be logged and filtered. 
Generated code that uses common LLM binaries or tools such as PowerShell should be blocked. 

\textit{Anomaly Detection: } Anomalies such as excessive requests to generate code/scripts, reconnaissance prompts, and unusual access times or volumes should trigger alerts. 

\textit{Tool Use Restrictions: } As LLMs become more agentic and use tools on-device, restrict LLMs to only those tools that are necessary. 

\textit{LLM Usage Restrictions: } Allow users to disable the use of LLMs for code generation if they do not need it for that purpose, which will severely restrict the ways that threat actors can use the tool. 

\textit{Crowdsourced Rules for LLM Abuse Patterns: } Similar to the way that Snort rules can be used to detect network attacks, standard formats should be developed to detect LLM abuse patterns and crowdsourced to provide active threat intelligence. 

\vspace*{-8pt}

\section{ACKNOWLEDGMENTS}
Notice: This manuscript has been authored by UT-Battelle, LLC under Contract No. DE-AC05-00OR22725 with the U.S. Department of Energy. The United States Government retains and the publisher, by accepting the article for publication, acknowledges that the United States Government retains a non-exclusive, paid-up, irrevocable, world-wide license to publish or reproduce the published form of this manuscript, or allow others to do so, for United States Government purposes. The Department of Energy will provide public access to these results of federally sponsored research in accordance with the DOE Public Access Plan (http://energy.gov/downloads/doe-public-access-plan).

This manuscript was prepared as part of the Emerging and Cyber Security Technologies initiative at Oak Ridge National Laboratory. 

\def\refname{REFERENCES}

\bibliographystyle{plainnat}
\bibliography{references}

\begin{IEEEbiography}{Sean Oesch}{\,}is a researcher at Oak Ridge National Laboratory, Oak Ridge, Tennessee. His current research interests include autonomous cyber defense, explainable AI for cyber applications, and AI for cyber defense. Sean received his Ph.D. from the University of Tennessee, Knoxvile, where he studied password manager security. He is a Senior Member of IEEE. \vspace*{8pt}
\end{IEEEbiography}

\begin{IEEEbiography}{Jack Hutchins}{\,} is a researcher at Oak Ridge National Laboratory, Oak Ridge, Tennessee. His currently interest include AI security, LLMs, and robotics. Jack is currently pursuing a PhD at the University of Tennessee, Knoxville, where his focus is on robust artificial intelligence.\vspace*{8pt}
\end{IEEEbiography}

\begin{IEEEbiography}{Kevin Kurian}{\,} is a graduate student intern at Oak Ridge National Laboratory, Oak Ridge, Tennessee. His currently interest include LLMs and multiagent systems. Kevin is currently pursuing a PhD at the University of Florida\vspace*{8pt}
\end{IEEEbiography}

\vspace*{8pt}
\begin{IEEEbiography}{Luke Koch}{\,} is a counter-AI researcher at Oak Ridge National Laboratory, Oak Ridge, Tennessee. His current research interests include demonstrating malicious machine learning models and enhancing security for HPC environments. He received his PhD from the University of Tennessee, Knoxville, where he focused on countering AI-based malware detection via static binary instrumentation.\vspace*{8pt}
\end{IEEEbiography}
\end{document}